\documentclass[
 reprint,groupedaddress,showpacs,amsmath,amssymb, aps,pra]{revtex4-1}

\usepackage{graphicx}
\usepackage{dcolumn}
\usepackage{bm}

\begin{document}

\title{Perturbation Theory for Arbitrary Coupling Strength?}

\author{Bimal P. Mahapatra$^{1}$}
\email{bimal58.mahapatra@gmail.com}
\altaffiliation[Also formerly, at ]{School of Physics, Sambalpur University, Jyoti Vihar-768019, India.}
\author{Noubihary Pradhan$^2$}
\email{noubiharypradhan1963@gmail.com}
\affiliation{$^1$School of Physical Sciences, National Institute of Science Education and Research (NISER), Bhubaneswar 751005, India\\$^2$ Department of Physics, G.M.University, Sambalpur 768004, India}
\date{\today}

\begin{abstract}
We present a \emph{new} formulation of perturbation theory for quantum systems, designated here as: `mean field perturbation theory'(MFPT), which is free from power-series-expansion in any physical parameter, including the coupling strength. Its application is thereby extended to deal with interactions of \textit{arbitrary} strength and to compute system-properties having non-analytic dependence on the coupling, thus overcoming the primary limitations of the `standard formulation of perturbation theory' ( SFPT). MFPT is defined by developing perturbation about a chosen input Hamiltonian, which is exactly solvable but which acquires the non-linearity and the analytic structure~(in the coupling-strength)~of the original interaction through a self-consistent, feedback mechanism. We demonstrate Borel-summability of MFPT for the case of the quartic- and sextic-anharmonic oscillators and the quartic double-well oscillator (QDWO) by obtaining uniformly accurate results for the ground state of the above systems for arbitrary physical values of the coupling strength. The results obtained for the QDWO may be of particular significance since `renormalon'-free, unambiguous results are achieved for its spectrum in contrast to the well-known failure of SFPT in this case.
  

\end{abstract}

\maketitle

1.\emph{Introduction.}

 Approximation methods are inevitably employed~\cite{ar1} in the investigation of interacting quantum systems since exact analytical results are sparse. Perturbation theory continues to be the preferred method of approximation for various practical and theoretical reasons \cite{ar2}. However, it is well known that the standard formulation of perturbation theory (SFPT) suffers from a number of limitations arising primarily from its {\em defining} property of small-coupling-power-series-expansion. Perhaps the most severe limitation of SFPT is its inability to describe the important class of the so-called `non-perturbative' phenomena governed by large values of the coupling-strength and/or non-analytic-dependence on it. Other problems in SFPT include: the instability of the `perturbative-vacuum'~\cite{ar3} and inconsistency with the known analytic properties \cite{ar4} of physical observables as a function of the coupling-strength.
 
 Even for small values of the coupling, the perturbation-series(PS) is generically \cite{ar5} divergent but asymptotic in nature:$~|\mathcal{E}_k|\sim c^k \Gamma(ak+b)$, for $k\gg1$, where $a,b,c$ are constants depending upon the particular theory under consideration; $\mathcal{E}_k$ is the `$k$-th'-order contribution in SFPT defined by $E(g) = \mathcal{E}_0+\sum\limits_{k=1}^{\infty}g^{k}\mathcal{E}_k\equiv \mathcal{E}_0+\Delta\mathcal{E}(g)$ and $E(g)$ is an observable of the system . The asymptotic-nature of the PS manifests through the following property \cite{ar6}: ~~~~~~~~~~~~~~~~~~~~~~~~~~~~~~~~~~~~~~~~~~~~~~~~$\lim_{g \rightarrow 0_+}(~E(g)-\sum\limits_{k=0}^{N}g^{k}\mathcal{E}_k~)/~g^{N+1}=\mathcal{E}_{N+1}$. The standard procedure for constructing the analytic function $E(g)$ from its divergent, asymptotic series about $g\rightarrow{0_+}$  is via the method of Borel-summation \cite{ar5,ar6}. Indeed, the `sum' of the divergent, asymptotic PS is now customarily \textit{defined} \cite{ar5,ar6} by its Borel-sum when the latter exists.
 
 However, it is well known that Borel-summability of the perturbation series {\em fails} \cite{ar7} for \textit{any} value of the coupling-strength in case of several important physical systems characterized by degenerate  ( or, nearly degenerate ) ground state such as the quartic double-well oscillator(QDWO), QCD, the Fokker-Planck-system etc. This situation, thus further compounds the already stated limitations of SFPT.
 
 In view of the above mentioned inadequacies of SFPT, it is perhaps imperative to explore whether perturbation theory could at all be liberated from the limitations of small-coupling-power-series-expansion. In this note, we carry out such an exploration in some detail starting with one dimensional systems with anharmonic-interactions in quantum mechanics (QM). These systems, apart from their intrinsic merit due to numerous applications in diverse areas of physics, have also been traditionally employed (see, e.g.\cite{ar2}) as `theoretical laboratory' for testing new ideas in approximation schemes. Additionally, their significance in providing insight for the study of quantum field theory (QFT), have been known since a long time \cite{ar8, ar9}. Indeed,systematic investigations of large-order behavior of (renormalized) perturbation expansion in QFT \cite{ar10} were primarily motivated by similar studies \cite{ar8, ar9} in the above examples in QM. 
   
 In what follows, we choose the stationary-energy, $E_n(g)$ of the quantum-system as the observable, being defined by the eigen-value equation for the Hamiltonian: ${H}(g)|\psi_n(g)\rangle=E_n(g)|\psi_n(g)\rangle$. The Hamiltonian, $H(g)$ can be split into an \emph{exactly solvable} dominant part~$H_0$ and a \emph{sub-dominant} perturbation~$H'$ i.e. $H(g)~=~H_0~+~H'$.  For the realization of a perturbative-framework for arbitrary values of the coupling $g$, it may be \textit{sufficient} to fulfill the following two conditions:\\
  \textbf{(a)} both $H_0$ and $H'$ must depend \textit{non-trivially} upon the coupling strength $g$, i.e. $H_0=H_0(g)$ and $H'=H'(g)$ such that the dominant $g$-dependence resides in $H_0(g)$,\\
 \textbf{(b}) the contribution of the perturbing Hamiltonian $H'(g)$ remains sub-dominant for arbitrary value of $g$.
 
  A practical implementation of the above conditions can be achieved \cite{ar11,ar12} through the following steps:\\ 
   \textbf{(i)} choose $H_0$ to depend upon a suitable set of free parameters $\{{\alpha_i\}}$ i.e. $ H_0= H_0(\{\alpha_i\})$,\\ 
     \textbf{(ii)} impose the constraint that
  \begin{equation}
  \langle\phi_n|H(g)|\phi_n\rangle~=~\langle\phi_n|H_0(\{\alpha_i\})|\phi_n\rangle,
  \end{equation}
  for all physical `$n$' and `$g$' where, $|\phi_n\rangle$ are defined by the eigen-value equation for $H_0$: $H_0|\phi_n\rangle=E_0^{n}|\phi_n\rangle $, with  `$n$' denoting the spectral-label,\\
   \textbf{(iii)} determine the parameters $\{\alpha_i\}$ by the constraint given by (1) and by further variational-minimization of $\langle\phi_n|H(g)|\phi_n\rangle$, if needed.
  
   By this procedure, the non-linear $g$-dependence of the system-Hamiltonian, $H(g)$ is effectively fed- back into the ``mean-field Hamiltonian(MFH)", $H_0$: 
  \begin{equation}
  H_0(\{\alpha_i\})\rightarrow ~H_0(g,n)
  \end{equation}

  The \emph{exact} solution of the eigen-value equation for  $H_0(g,n)$ given by $E_0^n(g)$, are then naturally identified as the leading-order (LO) result and these are expected to provide the major contribution for the true eigenvalue, $E_n(g)$ . This expectation has been explicitly verified~\cite{ar11,ar12} for the case of anharmonic-interactions (AHI) in one-dimension, by obtaining uniformly accurate results for $E_n(g)$ in the LO, for arbitrary physical values of `$g$'~and `$n$'.
     
  In view of (1), we designate this scheme as the ``mean-field approximation scheme (MFAS)" in analogy with the well-known method used in many-body-physics. 
  
  In the exact form of eq.(1), the approximation was applied to AHI in \cite{ar13}. In ref.\cite{ar14}, the same equation,eq.(1) was used under the name of ``generalized Hartree-approximation" in the study of the spectrum of anharmonic- and double-well oscillators and in exploring the structure and stability of the resultant ground state of the interacting system. 
  
  In this investigation, we employ the above scheme to formulate a perturbation theory with the objective of universal application to interactions of arbitrary strength. This is described in the next section.
  
  2.\emph{Mean field perturbation theory.}
  
  A new formulation of perturbation expansion, which we designate as ``Mean-Field-Perturbation Theory (MFPT)", is naturally suggested in view of the above considerations by defining the perturbation about the mean-field Hamiltonian as: $H'(g,n)\equiv H(g)-H_0(g,n)$. An immediate consequence of this prescription is the following result:
 \begin {equation}
  \langle\phi_n|H'(g,n)|\phi_n\rangle=0
  \end {equation} 
 for all `$n$' and `$g$', which follows from (1-2). This result ensures condition-\textbf{(b)} mentioned above (in the sense of quantum-average) since $\langle H'(g)\rangle\ll\langle H_0(g)\rangle\equiv E_0^n(g)$.
  (Here, the quantum-average of an operator $A$ is defined as: $\langle A \rangle\equiv\langle\phi_n|A|\phi_n\rangle$). Moreover, eq.(3) has a direct consequence that the first-order perturbation correction in the Rayleigh-Schr\"{o}dinger-perturbation series(RSPS) {\it vanishes} identically for all `$n$' and `$g$':
  \begin{equation}
  E_1^n\equiv\langle\phi_n|H^{\prime}|\phi_n\rangle= 0.
  \end{equation}
  
  As discussed later, these  properties: eqs.(1-3) have important bearing on the nature of the resultant perturbation-series in MFPT, which is defined below 
  (henceforth, we do not display the $n$-dependence of the various quantities, for notational-convenience):
  \begin{equation}
   E(g)= E_0(g)+\sum\limits_{k=1}^{\infty}{E_k(g)}\equiv {E_0(g)}+\Delta{E(g)}.
 \end{equation}
 For the study of convergence-properties of the above series, one needs to compute the energy-corrections  $E_k(g)$ to arbitrary order, $`k'$. However, since the above series is \emph{not} a power-series-expansion in $g$, we resort to the well-known \cite {ar15} recipe of introducing an auxiliary, \textit{dummy} parameter denoted as $\eta$, for generating a power-series in this parameter (chosen real) and project out the `$k$-th' order correction, $E_k(g)$ by the following procedure: consider an associated-Hamiltonian(AH), $\bar{H}$ given by $\bar{H}\equiv H_0+{\eta}H^{\prime}$, and the corresponding eigen-value equation:
  
   $~~~~~~~~~~~~\bar{H}(\eta,g)|\bar{\psi}(\eta,g)\rangle=\bar{E}(\eta,g)|\bar{\psi}(\eta,g)\rangle$,\\\\ such that $\bar{E}(\eta,g)$ can be expanded as a \emph{formal} power-series in $\eta$:\\
    $~~~~~~~~~~~~~~~~~\bar{E}(\eta,g)=\sum\limits_{k=0}^{\infty}{\eta^k}{E_k(g)}$.\\  The `$k$-th' perturbation-correction $E_k(g)$  being the same as appears in eq. (5), can then be identified from the above equation as the coefficient of ${\eta^k}$ \emph{before} setting the limit, $\eta\rightarrow1$. 
   
   It may be emphasized that the above procedure is merely an intermediate book-keeping device to project out the $E_k(g)$ appearing in eq.(5)-apart from achieving this goal, the above formalism plays no other role here. In particular, the final results are \emph{independent} of the dummy-variable $\eta$ since, by construction, $H(g)=\bar{H}(\eta=1,g)$, $E(g)=\bar{E}(\eta=1,g)$ etc.
        
  We determine the corrections $E_k(g)$ to arbitrary order $`k'$ in MFPT,  using the recursion relations derived from the application of the `hyper-virial theorem(HVT)' and the `Feynman-Hellman theorem(FHT)'. To this end, consider a non-relativistic one-dimensional system in quantum mechanics described by the Hamiltonian:\\
   $~~~~~~~~~~~~~~~~~~~~H= (1/2) p^2+V(x)$.\\ Then application of the HVT to this Hamiltonian leads to the following equation \cite{ar2}:
   \begin{equation}
   2E \langle{f^\prime}\rangle - 2 \langle{f^\prime}V \rangle-\langle f{V^\prime}\rangle
   +\left(\frac{1}{4}\right)\langle{f^{\prime \prime \prime}}\rangle~=~0,
   \end{equation}
   where the notation is as follows: $E =~$energy eigen-value; $f=f(x)$ is an arbitrary differentiable function which can be conveniently chosen for the particular problem at hand, `prime(s)' denote differentiation and  $\langle A\rangle\equiv\langle\psi|A|\psi\rangle $ for an operator $A$, with $|\psi\rangle$ denoting the (normalized) eigen-function of~$H$.\\ For the same system, the statement of the FHT reduces to:
   \begin{equation}
   \langle\psi|\frac {\partial {H}}{\partial\lambda}|\psi\rangle=\frac{\partial E}{\partial\lambda}
   \end{equation}
  Here, $\lambda$ generically represents any parameter, on which the system-Hamiltonian $H$ depends. (Typically, $\lambda$ could represent: `mass',`charge',`coupling-strength(s)', or even a \emph{dummy} variable introduced to define an \emph {auxiliary} expansion-parameter for perturbation theory).
                
  In the next section, we illustrate the implementation of the scheme as outlined above, by applying the same to anharmonic-interactions in one-dimension.
       
3(a).\emph{Application to the Quartic- and Sextic- Anharmonic Oscillators.}

 The Hamiltonian for both the cases are given by:
 \begin {equation}
  H~=~\frac{1}{2}p^{2} + \frac{1}{2}x^{2} + gx^{2K},~~\mbox(K=2,3),
 \end {equation}
  corresponding to the quartic-AHO (QAHO) and the sextic-AHO (SAHO) respectively. These quantum-systems are quite basic in view of their wide applications \cite{ar1,ar2} in diverse areas of physics and , as noted earlier, for their use as `theoretical-laboratories' in testing various aspects of approximation methods \cite{ar1, ar2},\cite{ar5,ar6,ar7},\cite{ar11,ar12}. In order to apply MFPT, we choose \cite{ar11} the `harmonic-approximation' for the input-MFH: 
\begin{align}
H_0~=~\frac{1}{2}p^{2} + \frac{1}{2}\omega^{2}x^{2} + h_{0},
\end{align}

The parameters, $\omega$ and $h_0$ are determined \cite{ar11} by following the steps \textbf{(i)-(iii)} outlined in Section-1. For the case of the QAHO, this leads \cite{ar11} to their determination as follows: $\omega$ is obtained as the real,+ve root of the equation:
\begin{align}
 ~\omega^{3}-\omega-6g(\xi+\frac{1}{4\xi})= 0,
 \end{align}
 and $h_0$ is given by  
\begin{align} 
 ~h_0=\left(\frac{\xi}{4}\right)\left(\frac{1}{\omega}-\omega\right)
\end{align}
where, $\xi\equiv(n+\frac{1}{2})$; with the spectral-index $n$ taking values: ~$ n=0,1,2,3,...$.
 (In the following, we will refer eq.(10) as the `gap-equation' and eq.(11) as the `energy-shift') 
 
 As expected, these parameters thus acquire the functional dependence on $g$ and $\xi$:
 \begin{subequations}
 \begin {align}
 \omega\rightarrow\omega(g,\xi),\\
 h_0\rightarrow h_0(g,\xi). 
 \end {align}
\end{subequations}
Similarly, the eigen-values of $H_0$ also acquire the required  $g,\xi$-dependence and given by~\cite{ar11}:
 \begin{align}
  ~E_0(g,\xi)=\left(\xi/4\right)\left(3\omega+\frac{1}{\omega}\right).
  \end{align}
 In view of the central importance of eqs.(10-13) in obtaining the results to follow, we sketch here the key-steps leading to these equations for sake of clarity and completeness and refer the reader to ref.[1,11] for further details.
To start with, one needs to evaluate:\\
$~~~~~~~~~~~~\langle H \rangle~=~(1/2)\langle p^2 \rangle + (1/2)\langle x^2 \rangle + g\langle x^4 \rangle $, \\ where now the notation is 
$\langle A\rangle\equiv\langle\phi_n|A|\phi_n\rangle $, for an operator $A$. Using the definition of $|\phi_n\rangle$ as the eigen-function of $H_0$, the relevant operator-averages occurring in $\langle H \rangle$ can be evaluated [1,11] by standard methods , e.g. by using the formalism of the ladder-operators for $H_0$ and are given by the following equations :  
\begin{subequations}
    \begin{align}
        <x^{2}> = (\xi / \omega) ,~~
        ~< p^{2}>=\omega \xi,\\
        ~~~<x^{4}>~ = (\xi / \omega) + 3~( 1 + 4~\xi^{2} )/8 \omega^{2} ,\\< x^{6} >~ =~( \xi/\omega ) 
        + (5/8) (\xi/\omega^{3})(5 + 4\xi^{2}).
       \end{align}
   \end{subequations}
   Hence, $<H>$ is evaluated as:
   \begin{equation}
   <\phi_n|H|\phi_n> = \omega\xi/2 +(\xi/2
   \omega)
   + (3g /8\omega^{2})(1+4\xi^{2}).
   \end{equation}
   Variational minimization  of eq.(15) with respect to `$\omega$' then leads to the `gap-equation', eq.(10). From eq.(1) and eq.(15), it then follows that the leading order(LO) result for the energy is given by eq.(13). Similarly, the expression for $h_0$ as given by eq.(11), follows by noting that $h_0=E_0-\omega \xi$. 
   
  At this point, several features/aspects regarding the leading order (LO) results based on eqs.(10-13) may be noted:\\
  (a) eq.(10) has been derived independently by several authors \cite {ar16} starting from widely different considerations.\\
   (b) Moreover, the rigorously established \cite{ar4} analytic-structure of $E(g)$ in the $g$-plane, as well as, the non-analytic dependence on the coupling strength`$g$', are trivially contained \cite{ar1,ar11} already at the LO-level through the solution of eq.(10), which is explicitly given by:
    \begin{equation}
    \omega~=~(3g f(\xi))^{1/3}[(~1 + \sqrt {(1 - \rho)})^{1/3}~+~(~1 - \sqrt 
    {(1 - \rho)})^{1/3}],
    \end{equation}
    where, $~\rho^{-1}~=~243g^ {2}f^{2}(\xi)~$ and $f(\xi)\equiv\xi+(1/4\xi)$.\\\\
    (c) The dependence of $\omega$ and $h_0$ on $\xi$ and $g$ implies that the `mean-field Hamiltonian', $H_0$ also depends on the same parameters. The physical significance of such dependence is that the eigen-functions of $H_0$, $|\phi_n\rangle$ are not mutually orthogonal: $\langle\phi_m|\phi_n\rangle\neq 0 $ for $m \neq n $. (As discussed later, this situation does \emph {not} cause any problem in the development of the Rayleigh-Schr\"{o}dinger-perturbation series (RSPS), which can be generated by a `wave-function-independent method', such as the HVT-FHT-based formalism employed here.)\\(d) A further significance of $\omega$ arises from the altered ground-state structure \cite {ar14} due to interaction, differing non-trivially from the free-field ground state. Also, the ``dressed ground state'' of the interacting system in the mean-field approximation has lower energy as compared to the ``trivial" ground state of the free-theory for \emph {any} non-vanishing value of the coupling $g$, thus establishing instability  of the free-field ground state in presence of interaction. This aspect has been discussed in detail elsewhere \cite {ar14} \\(e)The($g$,$\xi$)-dependence of $\omega$ and $h_0$ as determined in the LO, does \emph {not} get altered later in computing the perturbation corrections at higher orders. We elaborate on this important feature later (Section-4).\\ (f) Further, as stated earlier, the accuracy of the energy-spectrum obtained in the LO, eq.(13) is quite significant- the deviations from the `exact'-results being no more than a few-percent \cite{ar11} over the full-range of `$g$' and `$n$'. This result ensures that the dominant contribution comes indeed from the LO, as required in a perturbative frame-work.
  
We next turn to the development of the MFPT for the AHO's. The perturbation about the mean-field Hamiltonian is given by: $H^{\prime}~\equiv~H-H_0=gx^{2K} - \frac{1}{2}(\omega^{2} - 1)x^{2}-h_{0}$ and consequently, the associated Hamiltonian is given by: $\bar{H}=\frac{1}{2}p^{2} + \frac{1}{2}\omega^{2}x^{2} + h_{0} + \eta(g x^{2K} - \frac{1}{2}(\omega^{2} - 1)x^{2}-h_{0})$. 

 To apply the HVT to  $\bar{H}$, we make use of eq. (6); choose: $f(x)= x^ {2j};j= 0,1,2,3...$ and follow  analogous procedure as applied in ref. [2] to the case of the QAHO in SFPT, but now generalized to MFPT. This results in the following recursion-relation:
\begin{align}
X(j,i)=a(j)X(j-1,i)+ b(j)\sum_{m=0}^{i}E_{m}X(j-1,i-m)\nonumber\\
+c(j)X(j-2,i)-a(j)X(j-1,i-1)~~~~~~~~~~~\nonumber\\
+e_{\omega}X(j,i-1)-f(j)X(j+K-1,i-1),~~~~~
\end{align} 
where, $X(j,i)$ are defined by $X(j,\eta)\equiv~\langle\bar{\psi}|x^{2j}|\bar{\psi}\rangle=\sum\limits_{i=0}^{\infty}{\eta^i}X(j,i)$, with $\bar{\psi}$ and $\bar{E}$ as defined earlier and other quantities are given as follows:
\begin{align}
a(j)=(-h_{0}/\omega^{2})(2j-1)/j,~~ b(j)=(2j-1)/(\omega^{2}j)\nonumber,\\~~~~~~~~~~~~~~c(j)=~(j-1)[4(j-1)^{2}-1]/
(4j\omega^{2}),~~\nonumber\\~~~~e_{\omega}=(\omega^{2}-1)/\omega^{2},~~f(j)=(g/\omega^{2})(2j-1+K)/j.
\end{align}
Note that the input parameters in eq.(18), i.e. $h_0$ and $\omega$ are as determined in the LO, (See, eqs.(10,11)). To apply the FHT, we treat $\eta$ in $\bar {H}(\eta,g)$ as the variable on which the latter depends. This yields: 
$~~~~~~~~~~\langle\bar{\psi|}\dfrac {\partial\bar{H}}{\partial\eta}|\bar{\psi}\rangle=\dfrac{\partial\bar{E}}{\partial\eta}.~$\\
This equation provides additional relations between $X(j,i)$ and $E_m$, given by
\begin{subequations}
    \begin{align}
    E_{1}=gX(K,0)-\frac{1}{2}(\omega^2-1)X(1,0)-h_0,~~~~~~~~~~~~\\
    pE_{p}=gX(K,p-1)-\frac{1}{2}(\omega^2-1)X(1,p-1); p=2,3,4,...
    \end{align}
\end{subequations} 
The above eqs.(17-19) suffice to determine \textit{exactly}, the energy-corrections $E_p$  to \textit{arbitrary}-order $p$ when supplemented by the `boundary-conditions' \cite{ar2}: $X(0,i)=\delta_{0i}$, and $X(j,i)=0$ for $j<0$, where $\delta_{mn}$ denotes the Kr\"{o}necker-symbol. Note that the evaluation of $E_p$ requires the computation of $X(j,i)$ for the values of $i$,$j$ lying in the following range (in steps of unity): $~0\leq~i~\leq (p-1)$, and$~1\leq~j~\leq (K-1)(p-i)+1$. 

For the case of the QAHO ( i.e. by substituting $K$= 2 in eq.(17-19)), we record here the following results for $E_p(g,\xi)$ at $g=1$ and $\xi=\frac{1}{2}$ ( i.e.for the ground-state), for some low-lying values of $p$: $E_1=0, E_2= -\frac{3}{256}, E_3= \frac{27}{4096}, E_4= -\frac{2373}{262144}, E_5 = \frac{65457}{4194304}....$ etc.

It may be pertinent at this point to discuss the merits of the above formalism based on HVT-FHT. This method has several distinct advantages over other approaches: it produces  \textit{exact} analytic results for perturbation-corrections to an arbitrary order, whereas other methods are almost always approximate, being afflicted by unavoidable errors arising from the neglect of sub-asymptotic corrections, truncation-errors etc. Moreover, the recursive evaluation of the perturbation corrections yields to easy computation, which is an additional practical-advantage. Note also that the HVT-FHT-  method  does \emph{not} require \cite{ar2} the computation of eigen-functions,~$|\phi_n\rangle$ of $H_{0}$ or perturbation-corrections there to. This feature is of special advantage here since, as stated earlier, eq.(2) implies the non-orthogonality of the eigen-functions,~$|\phi_n\rangle$, which would otherwise prevent the use of the standard `text-book method'( see, e.g. ref. [15]) of derivation of the RSPS.

We next turn to the case of the sextic anharmonic oscillator (SAHO). Results for the SAHO can be obtained by following analogous procedure as in the case of QAHO but now substituting $K=3$ in eqs.(17-19), along with the following input \cite{ar11,ar1} for $\omega$, $h_0$ and $E_0$: $\omega$ is given by the real, positive root of the equation,\\ $~~~~~~~~~~~~~~~~~~\omega^4-\omega^2-(15g/4)(5+4\xi^2)=0$;\\ $~~~~~E_0=(\xi/3)(2\omega+\dfrac{1}{\omega})~~~~$    and  $~~~h_0=(\xi/3)(\dfrac{1}{\omega}-\omega)$.\\
(Again, the LO-results for energy are accurate \cite {ar1,ar11} to within a few percent over a wide range in $\xi$ and $g$ ). \\
 Sample-values of $E_p$ computed for the ground state of the SAHO and for $\omega=2$ are the following: $E_1=0, E_2= -\frac{49}{960}, E_3= \frac{671}{4608}, E_4= -\frac{53621891}{55296000}, E_5= \frac{2610955409}{265420800}....$ etc. As expected ~\cite{ar17}, the resultant asymptotic series exhibits more severe divergence at large orders than that of the QAHO.

 We consider next, the case of the quartic double-well oscillator.
 
3(b).\emph{The Quartic Double well Oscillator (QDWO).}

As already mentioned, the case of the QDWO is not Borel-summable \cite{ar7} in SFPT for any value of the coupling-strength. Several modern developments \cite{ar18,ar19} such as the theory of resurgence and trans-series\cite{ar18}, distributional-Borel-summation \cite{ar19} etc are basically motivated to surmount this  problem. In view of the above scenario, the case of the QDWO assumes special relevance for investigation in MFPT. 

The Hamiltonian in this case, is given by:

~~~~~~~~~~~$~H=(1/2)p^{2} - (1/2)x^{2} + gx^4,~~~~~~~~~~~~~~~~~$\\ 
  and the MFH is again chosen in the harmonic-approximation as:
 
  ~~~~~~~$H_0=(1/2)p^{2}+(1/2)\omega^{2}(x-\sigma)^2 + h_0$,\\
     but generalized to take into account, the spontaneous symmetry breaking (SSB), through a non-zero vacuum-expectation-value for $x$ denoted as $\sigma$. Accordingly, the various average values, analogous to eq.(14) are now given by:
  \begin{subequations}
       \begin{align}
   <~x^{2}~>~ =~ \sigma^{2} + (\xi / \omega)~ ,~< p^{2} >~=~\omega \xi,~~~~~~\\
   <~x^{3}~>~ =~ \sigma^{3} + 3\sigma (\xi / \omega),~~~~~~~~~~~~~~\\
   <~x^{4}~>~ =~ \sigma^{4} + 6~\sigma^{2}~(\xi / \omega)
    + 3~( 1 + 4~\xi^{2} )/8 \omega^{2}, 
         \end{align}
\end{subequations}

   Eqs.(20) enables the valuation of $<H>$ in terms of the input parameters . These parameters: $\omega$, $\sigma$ and $h_0$ are then determined analogously as in the case of the AHO, in terms of $g$ and $\xi$. However, a distinct feature in the case of the QDWO is the occurrence of `quantum-phase transition(QPT)' \cite {ar11} governed by a `critical-coupling' $g_c(\xi)$ given by the expression: $g_c(\xi)=\dfrac{(2/3)^{3/2}}{3(5\xi-(1/4\xi))}$, such that the SSB-phase is realized with $\sigma\neq 0$ for $g\leq g_c(\xi)$,  whereas the `symmetry-restored(SR)-phase' is obtained with $\sigma=0$ when $g > g_c(\xi)$. (Numerically, we have: $g_c=0.09718 $, for the ground-state of the DWO.) 
   
   The transition across $g=g_c(\xi)$ being discontinuous, the two phases are governed by distinct expressions for $\omega$ and $E_0$, which are \textit{not} analytically connected. It is, therefore, necessary to consider the two-phases separately. However, owing to the rather small-value of $g_c$ the `SSB-phase' exists only over a very limited range of $g$, i.e. when $0\leq g_c\leq 0.09718$ for the ground state and $g_c$ takes even smaller values for the higher excited states, eventually vanishing for large-$n$. We therefore, confine here to reporting the results for the `SR-phase' only. 

In this phase we have: $ g > g_c(\xi)$; $\sigma=0$; $\omega$ satisfies the equation,$~\omega^{3}+\omega-6g(\xi+\dfrac{1}{4\xi})= 0$; $E_0=(\xi/4)(3\omega-(1/\omega))$; $h_0= E_0-\omega\xi$. As in the case of the QAHO and the SAHO, the LO-results: $E_0$ capture the dominant contribution in reproducing the energy-spectrum [1,11] to within a few percent.\\ The expressions for $H^{\prime}$ and $\bar{H}$ are now respectively given by: $ H^{\prime}= gx^{4} - (1/2)(\omega^{2}+1)x^{2}-h_{0}$, $\bar{H}=(1/2)p^{2} + (1/2)\omega^{2}x^{2} + h_{0} + \eta(g x^{4} - (1/2)(\omega^{2} + 1)x^{2}-h_{0})$. The application of the HVT to $\bar{H}$ leads to the recursion relation \emph{identical} to that of QAHO as given by eq.(17-19) but with the following \textit{changes} in the definition of the input-parameters, $\omega$, $h_0$ and $E_0$. These are now as given above for the case of the SR-phase of the QDWO. Moreover, $K=2$ and $e_{\omega}$ (see, eqs.(17,18)) is now defined as: $~e_{\omega}=\omega{^2}/(\omega{^2}+1)$. Similarly, relations analogous to (19), follow from the application of FHT to this case: $~E_{1}=gX(2,0)-(1/2)(\omega^2+1)X(1,0)-h_0~$, $~pE_{p}=gX(2,p-1)-(1/2)(\omega^2+1)X(1,p-1); p=2,3,4,...$. Sample-values for$~E_p$ evaluated at $\omega=1$ and $\xi=1/2$ are the following: $E_1=0, E_2= -\frac{1}{24}, E_3=\frac{1}{16} ,E_4=-\frac{791}{3456}, E_5= \frac{7273}{6912}$....etc.\\\\
We next proceed to evaluate the total `sum' of the perturbation corrections. In that context, the following common-features that emerge from the  computations in \emph {all} the above cases may be noted: $E_1= 0$, and the perturbation-series have terms that \textit{alternate} in sign. These two features  in MFPT are found crucial in obtaining the `sum' of the perturbation-series as described below.
      
  4.\emph{Computation of the Total Perturbation Correction}\textrm{(TPC)}.
  
  We compute the TPC by two main methods:(a) the method of optimal truncation of the original asymptotic perturbation-series and (b) by Borel-summation. These are sequentially described below.(For convenience, we confine here to the computation for the \emph{ground-state}.)
  
  4.(a)\emph{Method of Optimal Truncation}.--
  This method is based on the property~\cite{ar6}, that the initial terms of an asymptotic-series continue to decrease in magnitude till the `term of the  least magnitude'(TLM) is reached beyond which, the subsequent terms exhibit monotonic increase. Therefore, the TPC can be reasonably approximated by truncation of the perturbation series at the TLM. In the case of MFPT, we have:\\ $~~~~~~~~~~~~~\left(E(g)-\sum\limits_{k=0}^{N}{E}_k(g)\right) \leq {{E}_{N+1}(g)}$.\\ Therefore, if $N_0(g)$ denotes the `order' of the TLM, the truncation of the series at term, $N_0(g)$ leads to the following estimate of the TPC: 
  \begin{equation}
  ~~E(g)~\simeq\sum\limits_{k=0}^{N_0(g)}{E}_k(g).
  \end{equation}
  In Table-I (columns 2-4), we present the results from this method for the case of the QAHO, QDWO and SAHO for sample-values of $g$ (column 1). It can be seen from this Table that the `method of optimal truncation' works rather efficiently for arbitrary physical value of $g$ and that the TLM occurs at fairly low values :
    $N_0(g)\sim~$ (2-6). The primary reason for this latter feature can be traced to the fact that $E_1(g)$ \textit{vanishes} for arbitrary $g$ (see, eq.(4)). Therefore, since the series must diverge at large orders, the TLM is constrained to occur at fairly low-values for \textit {all} values of $g$ without compromising the accuracy (Table-I).

        For comparison with `standard' results, we have considered earlier computations in \cite{ar16},\cite{ar20} and \cite{ar21} for the case of QAHO, SAHO and QDWO respectively. The relative-error/deviation from the standard results is seen (Table-I, column 4) to be within a few-percent, uniformly over the considered range of $g$. The `method of optimal truncation' (MOT) may, therefore, be regarded as a `fail-safe' method in MFPT for achieving reasonable accuracy over the full physical range of $g$.  This result could be of particular significance since the MOT is the only option left, when other methods such as  Borel-summation etc, fail or are inapplicable \cite{ar22} which, as stated earlier, occurs in several important cases of physical interest.

  4.(b){\em Method of Borel Summation}
  
   Consider the generic case when we have $E_j\sim~\Gamma(\alpha j+1)$ for $j\gg1$. Then, it follows that\\ $~~~~~~~~~~~~~~~~\lim_{j\gg 1}b_j\equiv \dfrac{E_j}{\Gamma(\alpha j+1)}\rightarrow 0.$\\ 
  Using the integral representation of the Gamma function: $\Gamma(z)=\int_0^{\infty}dt~exp(-t)t^{z-1}$, one can {\em formally} express the TPC as:
  \begin{align}
  \Delta{E(g)}\equiv\sum\limits_{k=1}^{\infty }E_k~\sim\gamma\int_0^{\infty}du~exp(-u^{\gamma})u^{\gamma-1}B(u),
  \end{align}
  where, we have defined, $u=t^{\alpha}$,$~\gamma=1/\alpha$ and $B(u)$ denotes the `Borel-Series' :  
  \begin{equation}
  B(u)\equiv\sum\limits_{j=1}^{\infty } b_ju^j. 
  \end{equation}
  Note   that,   although by   construction, $B(u)$ has a {\em finite} `radius of convergence'~$r_c$ , yet the `Borel-Laplace-integral', eq.(22) does \textit{not} exist as the range of integration extends beyond $r_c$. Denoting the \textit{analytic continuation} of $B(u)$ to the full-path of integration in (22) by $\tilde{B}(u)$ and substituting this for $B(u)$ there, the integral can be made to \textit{exist} and hence, can be used to\textit{`define'} the TPC as:
  \begin{equation}    
  \Delta{E(g)}\equiv\gamma\int_0^{\infty}du~exp(-u^{\gamma})u^{\gamma-1}\tilde{B}(u),
  \end{equation}
          
    We carry out the required analytic-continuation of $B(u)$ in eq.(23) by the use of conformal-mapping \cite{ar23}.
  This is the \textit{preferred} method \cite{ar23} when only a finite number of terms in eq.(23) are known/calculable.
  
   The input for constructing  a conformal map requires the knowledge of the nature and location of the singularity of $B(u)$, occurring closest to the origin in the $u$-plane. By the `Darboux theorem'\cite{ar24}, the late-terms in the series eq.(23) characterize this `leading'-singularity. This theorem is readily verified here considering the following \textit{ansatz} for the `leading-singularity-approximation' based upon the assumption of its `isolated' nature :\\ $~~~~~~~~~~~~~~~\tilde{B}(u)\simeq\tilde{B}_0(u)\sim~(u+r_c)^p$.\\ The `radius of convergence' $r_c$ and the `singularity-exponent'~$p$ can then be determined from a fast-converging set of equations \cite{ar2} given below:\\ $~~~~~~~~~r_c(g)=\lim_{j\gg1}\left(\dfrac{b_jb_{j-1}}{jb_j^2-(j+1)b_{j+1}b_{j-1}}\right),\\
  ~~~~~~~~~p(g)=\lim_{j\gg1}\left(\dfrac{{jb_j^2}-(j^2-1)b_{j-1}b_{j+1}}{jb_j^2-(j+1)b_{j+1}b_{j-1}}\right)$.\\
  We determine the value:$~p(g)=~-0.5$, being independent of $g$ as expected, when the relevant equation converges. Similarly, the values for $r_c(g)$ determined by convergence, are given in Table-I.  We find that these equations converge in all the cases investigated here, except for the case of the QDWO when the value of the coupling $g$ lies in the vicinity of the critical coupling $g_c$. In the event of non-convergence, we fix these parameters such that the Borel-integral converges ( see, later in this Section ). 
  
  Equipped with these inputs, we consider the conformal mapping $z$ \cite{ar23}, which maps the cut-u plane: $|arg(u)|<\pi$ into the interior of the unit-disk in the $z$-plane while preserving the origin:\\
  $~~~~~~~~~~~~~~~~~z(u)=\left(\dfrac{\sqrt{1+su}-1}{\sqrt{1+su}+1}\right)$,\\ where $s=(1/{r_c})$; $u=t^{\alpha}$.
  Note that the images of the points: $u=\pm\infty$ are at $z=1$ and further that $z(u)$ is a real-analytic-function of $u$. Moreover, the above mapping is proven to be ``optimal" by the {\it Ciulli-Fischer Theorem} \cite{ar25}, in the sense that the error of truncation becomes \textit{minimum} when an arbitrary function $f(z)$ expanded as an infinite- power-series in $z$  is approximated by truncation at a finite number of terms. Hence, using this property to advantage, we consider the partial sum of the Borel-Series as defined in eq.(23):\\ $~~~~~~~~~~~~~~~~~~~~~~B_N(u)\equiv\sum\limits_{k=1}^{N} b_ku^k$.\\ By the use of the inverse-transformation, $u(z)=\left(\dfrac{4}{s}\right)\dfrac{z}{(1-z)^2}$, one obtains:  $ u^N\sim~ z^N+O(z^{N+1})$ for $|z|< 1$. Hence, $B_N(u)$ can be consistently re-expanded \cite{ar26} as follows:
    \begin{subequations}
         \begin{align}
      B_N(u)\equiv\sum\limits_{k=1}^{N} b_ku^k\rightarrow~\bar{B}_N(z)=\sum\limits_{k=1}^{N} B_kz^k,\\
      \mbox{where,}~~ B_k=\sum\limits_{n=1}^{k}b_n\dfrac{\left(n+k-1\right)!}{\left(k-n\right)!\left(2n-1\right)!},
      \end{align}
    \end{subequations}
    
    \begin{widetext}
        
        \begin{table}
            
            \caption{Results for the total-perturbation-correction (TPC) to the ground state energy of the QAHO, SAHO and QDWO are presented for different values of the coupling-strength $g$ in col.(1). The legends used and entries in other columns are as follows: $E_{MOT}$ in col.(3) are the computed energy-values by the `method of optimal-truncation'(MOT) , see eq.(21); $N_0$ in col.(2) displays the value of the cut-off at the `term of least magnitude(TLM)' ; in col.(4)  the relative-( $\%$)error of approximation by $E_{MOT}$ with respect to entries shown in col.(10) marked as`Exact', are given ; in col.(5) $r_c$ are the input- values of the `radius of convergence' of the Borel-series defined in eq.(23); $N_c$ in col.(6) denotes the number of terms retained in the Borel-series, eq.(25a) for convergence up to the required accuracy; $\varDelta{E}$ in col.(7) represents the TPC due to the  Borel-sum in eq.(26a), after convergence; $E_0$ in col.(8) denotes the LO-contribution ; $E_{tot}$ in col.(9) represents the `total corrected energy' after Borel-summation; entries in col.(10) labeled $Exact$ display the `standard'-numerical results from the indicated reference and the last column specifies the ${\% error}$ of approximation by $E_{tot}$ with respect to the `Exact' result. For each case of anharmonic-interaction, the input-value of $\gamma$, (see eq.(24)) is indicated. We have fixed the value:~$\epsilon=0.001$ ,( see eq.(26a)) in the computation of TPC in all-cases.}
            
            \begin{tabular}{c c c c c c c c c c c}
                \hline
                $g$& $N_0$&$E_{MOT}$&$Er({\%})$&$r_c$&$N_c$& $\varDelta{E}$&$E_0$&$E_{tot}$& $Exact$&  $Er({\%})$\\
                \hline
                QAHO& &$(\gamma= 1)$ & &  & & & & & $ref.(16)$& \\
                
                \hline     
                0.1 & 6& 0.5593& 0.03 & 6.071  & 6 & -0.00116  & 0.5603   & 0.5591 &  0.5591 & 0.008 \\ 
                1.0 &3 & 0.8074& 0.44 & 2.667  & 7 & -0.00869  & 0.8125  & 0.8038&  0.8038 & 0.004\\ 
                10.0 &3 & 1.5204& 1.02 & 2.133  & 8 & -0.02619  & 1.5312  & 1.5050&    1.5050 & 0.002\\ 
                100.0 &3 & 3.1701& 1.23 & 2.028  & 10 & -0.06101 & 3.1924 & 1.1314 &   3.1314 & 0.0005\\ 
                \hline
                SAHO & & $(\gamma=1/2)$& & & & & &   &$ref.(20)$ &\\
                \hline
                0.1 &2 &  0.5787& 1.40 & 13.3 &  20& -0.0095 &  0.5964  & 0.5869 &   0.5869 & 0.001 \\ 
                1.0 & 2& 0.7694& 4.42 & 8.56 &  20 & -0.0328&  0.8378  & 0.8050 &   0.8050 & 0.002\\ 
                50.0 & 2& 1.7241& 7.23 & 7.14&  20 &-0.1149 &  1.9735  & 1.8586 &   1.8585  & 0.007\\ 
                200.0 & 2& 2.3986& 7.54 & 7.02& 20 &-0.1662 &  2.7606 & 2.5944 &   2.5942 & 0.007 \\ 
                \hline
                QDWO &  & $(\gamma = 0.8196)$& & & & &   & &$ref.(21)$ &\\
                \hline
                0.1& 2 & 0.4107 & 12.79 & 0.95 & 26 & -0.0787 & 0.5496 & 0.4709 & 0.4709 & 0.00006 \\
                \hline
                &  & $(\gamma =1)$& & & & &   & &     &\\
                \hline
                0.5 & 2& 0.4414 & 2.74 & 1.191 &  20&-0.0232 &  0.4770 & 0.4538 &    0.4538 & 0.0027 \\ 
                1.0 & 3& 0.5667 & 1.83 & 1.455 &  11 & -0.0216  & 0.5989& 0.5773 &   0.5773 & 0.0042\\ 
                10.0 & 3 & 1.4007& 1.66& 1.872 & 20 & -0.0320& 1.4098 & 1.3778 & 1.3778 & 0.0040\\ 
                100.0 & 3 & 3.1122& 1.37& 1.972 & 18& -0.0637& 3.1338 & 3.0701&   3.0701 & 0.0005\\ 
                \hline                
            \end{tabular}
            \label{$Table-I$}
        \end{table}
        
    \end{widetext}
    
    where, we have denoted the analytic-continuation of the partial sum of the Borel-Series by: $~B_N(u)\rightarrow~\bar{B}_N(z)$. Using the above properties/results, the `total-perturbative contribution' can be evaluated by the change of the integration-variable from $u$ to $z$ and substitution of $\bar{B}_N(z)$ for $\tilde B(u)$.  Note that in contrast to the Borel-series $B(u)$ in eq.(23), term-by-term integration of the partial-sum,$~\bar{B}_N(z)$ can now be carried out in eq.(25a) due to the incorporation of analyticity through the conformal-map $z$. Hence, considering the convergence in the limit of the sequence of partial-sums and by taking care of possible numerical-divergence that may occur at the upper limit, $~z=1$,  we finally obtain ( we suppress the $g$-dependence of quantities for notational-clarity ):
 \begin{subequations}                \begin{align}    
        \Delta{E}=\lim_{N\gg1,~\epsilon\rightarrow0}\left(\Delta{E}\right)_N\equiv\int_0^{1-\epsilon}dz f_N(z),\\
        \mbox{~with~}f_N(z)=(\gamma\rho)~\dfrac{(1+z)}{(1-z)^3}\left(\dfrac{\rho~z}{(1-z)^2}\right)^{\gamma-1}\nonumber\\~exp\left[-\left(\dfrac{\rho~z}{(1-z)^2}\right)^{\gamma}~\right]\bar{B}_N(z),
        \end{align} 
        \end{subequations}
        
      where,$~\rho=4r_c$ .
       The results are presented in Table-I. It may be seen from this Table that the convergence of the partial-sum is quite fast-- an accuracy of $\sim(99.999{\%})$ is uniformly achieved  over the full explored range of $g$ and for \textit{all} the  cases of anharmonic-interaction, when the cut-off $N_c$ on the terms retained for convergence of the partial-sum, is no more than $\sim~O\left(10-20\right)$ . Thus, the `$optimality$' and the `$accuracy$' of the method of conformal mapping, are firmly established in computing the total-perturbative contribution by the Borel-summation in MFPT.
     
     5.\emph{Discussion.}
     
     Before concluding, we discuss relevant investigations in the literature, which are close in spirit and content to MFPT. These investigations mostly make use of the combined techniques of variation- and perturbation methods and include the work of Caswell, ref.[16]; Killingbeck \cite{ar27}; and Kleinert and collaborators \cite{ar28}. The `$\delta$-expansion method', \cite{ar29}  and the algebraic techniques based upon `operator-expansion-method' \cite{ar30} also belong to this class. The approach based upon the `Gaussian effective potential(GEP)'\cite{ar9} is contained \cite{ar11} in MFPT in the `harmonic-approximation' at the leading-order. Hence, subsequent higher order corrections via MFPT correspond to the systematic perturbative improvement of the GEP.  
     
      The main feature of MFPT which distinguishes it from similar other approaches is that the defining equation, eq.(5) does not involve series-expansion in any physical-parameter. Further, it may be emphasized that there are no input parameters to be adjusted order-by-order in perturbation theory-- once they get determined at the leading order these remain unchanged later in computing corrections at higher orders. Some other distinct features of this method include: the simplicity of the scheme and  its potential for universal applicability to arbitrary Hamiltonian systems--the prescriptions outlined in Section-1 through steps: \textbf {(i)-(iii)} for implementation of MFPT are rather simple, straight-forward and general. Similarly , the generation of exact pertubation-corrections to arbitrary order through the recursion relations following from the combined use of the hyper virial- and Feynman-Hellman theorems, is also rather simple and general and leads to easy computation. The flexibility in the choice of the input Hamiltonian $H_0$, \cite {ar12} is another distinct feature, which may be used to advantage in improving the accuracy of approximation still further, as well as to test the stability of the scheme to different such choices. For the examples investigated here, the \emph {natural} incorporation of the analyticity (in the coupling-strength) and scaling properties of the system-observables to all orders has been demonstrated and we believe this feature to survive for other examples as well.  The vanishing of the first-order perturbation-correction, as implied by (1-2), is yet another characteristic feature of MFPT, which has important implication for computing the total perturbative-corrections.

    6.\emph{Summary, Conclusions and Outlook.}
    
    We have presented a new formulation of perturbation theory based upon a mean-field-like Hamiltonian-approximation, which does not involve power-series expansion in any \emph{physical} small-parameter including the coupling strength.
    Therefore, it is potentially applicable to interactions of arbitrary strength as well as, to compute system-properties involving non-analytic dependence on the coupling strength, thereby overcoming the primary limitations of the `standard formulation of perturbation theory' (SFPT).
    
     To test the scheme, we have applied it to compute corrections to the ground-state energy of `anharmonic-interactions' in one dimension, which have been traditionally considered (see,e.g.ref.[2]) as the bench-mark system to test new approximation methods. In this work, we have computed {\em Exact} perturbation-corrections to arbitrary order by the simultaneous use of the hyper-virial and Feynman-Hellmann theorems. Consequently, we have demonstrated that the resultant perturbation-series is Borel-summable for arbitrary physical value of the coupling-strength in the case of quartic- , sextic anharmonic oscillator and the quartic double-well oscillator. The perturbatively-corrected result for the energy of the ground state are shown to be uniformly accurate for sample values of the coupling-strength over the full physical range.
    
   The results for the quartic-double-well oscillator (QDWO) could be of  particular significance, since MFPT {\em naturally} leads to  perturbation-series, which is `renormalon'-free and   Borel-summable for arbitrary coupling-strength. This is to be contrasted with the situation in SFPT, which is {\em not} Borel-summable  for {\em any} value of the coupling  because of the occurrence of the `renormalon'-singularity in this case. As discussed earlier, this situation has primarily resulted in considerable efforts spent in recent literature \cite{ar7,ar18} to surmount this problem of SFPT for the QDWO and other systems, which are characterized by degeneracy (or near degeneracy) of the ground state.  
   
   On the basis of the results obtained here, it may be reasonably inferred that the distinction/divide between the `perturbative-' and the `non-perturbative'-regimes and the existence of other pathologies as mentioned earlier, could simply be {\em artifacts} of the SFPT.
        
   The work reported here can be extended in several directions. One immediate task could be to apply \cite{ar31} MFPT to other systems known to be Borel-non summable in SFPT to determine whether summability  can be achieved in such cases. Similarly, tunnel-splitting of energy levels can be studied \cite{ar31} after computation of perturbation correction to the excitation-spectrum of the QDWO using more realistic input-Hamiltonian, if necessary. Application to other systems are envisaged in a straight-forward manner in view of the universal nature of the approximation scheme.    
           
   \begin{acknowledgments}
     \emph{Acknowledgements.--}
     The major part of this investigation was carried out at the National Institute of Science Education and Research (NISER), Bhubaneswar, India while one of the authors (BPM) served there as a visiting Professor. He acknowledges the facilities for research extended to him at the Institute. The authors are grateful to Dr. F. M. Fern\'{a}ndez and to Mr. A. B. Mahapatra for help in Maple-programming and for computation, respectively and to the anonymous reviewer for helpful comments and suggestions.  
    \end{acknowledgments}

  \end{document}